\documentstyle [12pt]{article}
\begin{document}
\baselineskip .3in
\begin{titlepage}
\begin{center}{\large {\bf The travelling salesman problem on randomly diluted 
lattices: Results for small-size systems}}
\vskip .2in
{\bf Anirban Chakraborti}$~^{(1)}$ and
{\bf Bikas K. Chakrabarti}$~^{(2)}$\\
{\it Saha Institute of Nuclear Physics},\\
{\it 1/AF Bidhan Nagar, Calcutta 700 064, India.}\\
\end{center}
\vskip .3in
{\bf Abstract}\\
\noindent
If one places $N$ cities randomly on a lattice of size $L$,
we find that $\bar l_E\sqrt p$ and $\bar l_M\sqrt p$ vary with 
the city concentration $p=N/L^2$, where $\bar l_E$ is the average optimal travel 
distance per city in the Euclidean metric 
and $\bar l_M$ is the same in the Manhattan metric. We have studied such optimum
tours for visiting all the cities using a branch and bound algorithm, giving the
exact optimized tours for small system sizes ($N\leq 100$) and near-optimal
tours for bigger system sizes ($100<N\leq 256$). 
Extrapolating the results for $N\rightarrow \infty$, we find that $\bar l_E\sqrt
p = \bar l_M\sqrt p = 1$ for $p=1$, and $\bar l_E\sqrt p=0.73\pm 0.01$ and 
$\bar l_M\sqrt p=0.93\pm 0.02$ with $\bar l_M/\bar l_E \simeq 4/\pi$ as 
$p\rightarrow 0$. Although the problem is 
trivial for $p=1$, for $p\rightarrow 0$ it certainly reduces to the standard 
travelling salesman problem on continuum which is NP- hard.
We did not observe any irregular behaviour at any intermediate point.
The crossover from the triviality to the NP- hard 
problem presumably occurs at $p=1$.
\vskip 2.5in
\noindent
{\bf PACS No. :} 05.50+q
\end{titlepage}
\newpage
\noindent
{\bf 1 Introduction}\\
\noindent
The travelling salesman problem (TSP) is a simple example of a multivariable combinatorial 
optmization problem and perhaps the most famous one. Given a certain set of 
cities and the intercity distance metric, a travelling salesman must find the 
shortest tour in which he visits all the cities and comes back to his starting 
point. It is a non-deterministic polynomial complete (NP- complete) problem 
[1-3]. 
In the standard formulation of TSP, we have $N$ number of cities distributed 
randomly on a continuum plane and we determine the average optimal
 travel distance per city $\bar l_E$ in the Euclidean metric (with 
$\Delta r_E= \sqrt {\Delta x^2+\Delta y^2}$), or $\bar l_M$ in the Manhattan 
metric (with $\Delta r_M= |\Delta x|+|\Delta y|$).
Since the average distance per city scales (for fixed area) with the number of 
cities $N$ as $1/ \sqrt N$, we find that the normalized travel distance per 
city $\Omega_E=\bar l_E \sqrt N$ or
$\Omega_M=\bar l_M \sqrt N$ become the optimized constants and their values
depend on the method used to optimize the travel distance. Extending the 
analytic estimates of the average nearest neighbour distances, in particular
within a strip and varying the width of the strip to extremize (single parameter
optimization approximation), one gets $\frac {5}{8}<\Omega_E<0.92$ [4] and
$\frac {5}{2\pi}<\Omega_M<1.17$ [5]. Careful (scaling, etc.) analysis of the 
numerical results obtained so far indicates that $\Omega_E\simeq 0.72$ [6].

Similar to many of the 
statistical physics problems redefined on the lattices, e.g., the statistics of 
self-avoiding walks on lattices (for investigating the linear polymer 
conformational statistics), the TSP can also be 
defined on randomly dilute lattices. The (percolation) cluster statistics of 
such dilute lattices is now extensively studied [7].
The salesman's optimized path on a dilute lattice is necessarily a
 self-avoiding one; for optimized tour the salesman cannot afford to visit any
city more than once and obviously it is one where the path is non-intersecting.
The statistics of self-avoiding walks on dilute lattices has also been 
studied quite a bit (see e.g., [8]). However, this knowledge 
 is not sufficient to understand the TSP on similar lattices. The TSP on dilute
lattices is a very intriguing one, but has not been studied intensively so far.

The lattice version of the TSP was first studied by Chakrabarti [9].
In the lattice version of the TSP, the $N$ cities are represented by randomly 
occupied lattice sites of a two- dimensional square lattice ($L \times L$), 
the fraction of sites occupied being $p$ ($=N/L^2$, the lattice occupation 
concentration). One must then find the shortest tour in which the salesman 
visits each city only once and comes back to its starting point. The average 
optimal travel distance in the Euclidean metric 
$\bar l_E$, and in the Manhattan metric $\bar l_M$, are functions of the lattice
occupation concentration $p$ [10]. We intend to study here the variation of the 
normalised travel distance per city, 
$~\Omega_E=\bar l_E \sqrt p$ and $\Omega_M=\bar l_M \sqrt p$, 
with the lattice concentration $p$ for different system sizes.
It is obvious that at $p=1$, all the self-avoiding walks passing through all the
occupied sites will satisfy the requirements of TSP and $\Omega_E=1=\Omega_M$ 
(the distance between the neighbouring cities is equal to the unit lattice 
constant and the path between neighbouring sites makes discrete angles of 
of $\pi /2 $ or its multiples with the Cartesian axes). The problem becomes 
nontrivial as $p$ decreases from unity: isolated occupied cities and branching 
configurations of occupied cities are found here with finite probabilities and
self-avoiding walks through all the occupied cities, and only through the 
occupied cities, become impossible.
As $p$ decreases from unity, the discreteness of the distance of the path
connecting the two cities and of the angle which the path makes with the Cartesian
axes, tend to disappear. The problem reduces to the standard TSP on the 
continuum in the $p\rightarrow 0$ limit when all the continuous sets of 
distances and angles become possible. 
We study here the TSP on dilute lattice employing a
computer algorithm which gives the exact optimized tours for small system sizes
($N\leq 100$) and near-optimal tours for bigger system sizes ($100<N\leq 256$).
 Our study indeed indicates that $\Omega_E$ and $\Omega_M$ vary 
with $p$ and $\Omega_E\simeq 0.73$ and $\Omega_M\simeq 0.93$ as 
$p\rightarrow 0$. 

\noindent
{\bf 2 Computer Simulation and Results}\\
\noindent
We generate the randomly diluted lattice configurations following the standard
Monte Carlo procedure for different system sizes. For each system size $N$, we 
vary the lattice size $L$ so that the lattice concentration $p$ varies. 
For each such lattice configuration, the optimum tour with open boundary
conditions, is obtained with the help of the {\it GNU tsp\_ solve} [11] 
developed using a branch and bound algorithm (see Fig. 1). 
It claims to give exact results for $N\leq 100$ and near-optimal solutions for 
$100<N\leq 256$. It may be noted that the program works 
essentially with the Euclidean distance. However there exists a geometric 
relationship between the Euclidean distance and the Manhattan distance. We may
write $l_E=\sum_{i=1}^{N} r_i$, and $l_M=\sum_{i=1}^N r_i \alpha_i$, where $r_i$
is the magnitude of the Euclidean path vector between two neighbouring cities
and $r_i\alpha_i=r_i(|\sin \theta_i |+|\cos \theta_i |)$ is the sum of the components 
of the Euclidean path projected along the Cartesian axes. 
Naturally, $1\leq \alpha_i \leq \sqrt 2$. If $l_E$ corresponds to the shortest
Euclidean path, then $\sum_{i=1}^N r_i^{\prime} > \sum_{i=1}^N r_i$ , 
for any other path denoted by the primed set. If the optimized Euclidean
path does not correspond to the optimized Manhattan path, then one will have 
$\sum_{i=1}^N r_i^{\prime } \alpha_i^{\prime } < \sum_{i=1}^N r_i\alpha_i$,
where all the $\alpha_i$ and $\alpha_i^{\prime}$ satisfy the previous bounds.
Additionally, for random orientation of the Euclidean distance with respect to 
the Cartesian axes, $\langle \alpha_i \rangle =\langle \alpha_i^{\prime}\rangle=
(2/\pi) \int_0^{\pi /2} (\sin \theta + \cos \theta ) d\theta =4/\pi $. It seems,
with all these constraints on $\alpha$'s and $\alpha^{\prime}$'s, it would be 
impossible to satisfy the above inequalities on   
$\sum r_i$, and $\sum r_i \alpha_i$. In fact, we checked for a set of $50$ 
random optimized Euclidean tours for small $N$ ($<10$), obtained using the algorithm, whether the 
optimized Manhattan tours correspond to different sequence (of visiting the 
cities), and did not find any. We believe that the optimized Euclidean tour
necessarily corresponds to the optimized Manhattan tour.
We then calculate $l_E$ and $l_M$ for each such optimized tour.

At each lattice concentration $p$, we take about $100$
lattice configurations (about 150 configurations at some special points near 
$p\rightarrow 0$) and then obtain the averages $\bar l_E$ and 
$\bar l_M$. We then determine $~\Omega_E=\bar l_E ~\sqrt p~$ and
$~\Omega_M=\bar l_M ~\sqrt p~$ and study the variations of $\Omega_E$ and
$\Omega_M$, and of the ratio $\Omega_M /\Omega_E$ with $p$. 
We find that $\Omega_E$ and $\Omega_M$ both have variations starting 
from the exact result of unity for $p=1$ to the respective constants 
in the $p\rightarrow 0$ limit. In fact we noted that although $\Omega_M$ 
continuously decreases as $p\rightarrow 0$, it
remains close to unity for all values of $p$. 
We studied the numerical results for $N=~64,~81,~100,~121,~144,~169,
~196,~225~\rm {and}~256$. 
The results for $N=64~\rm {and}~100$ have been shown in Figs. 2 and 3
respectively. 
We have studied the variations in the values of $\Omega_E$ and $\Omega_M$ 
against $1/N$ for $p\rightarrow 0$, to extrapolate its value in the 
$N\rightarrow \infty$ limit. It appears that for the large $N$ limit (see 
Fig. 4), $\Omega_E(p\rightarrow 0)$ and $\Omega_M(p\rightarrow 0)$ eventually 
extrapolate to $0.73\pm 0.01$ (as in 
continuum TSP) and to $0.93\pm 0.02$, respectively. 
This result for $\Omega_E$ (at $p\rightarrow 0$) compares very well with the 
previous estimates [6].   As $p$ changes from $1$ to $0$, the ratio $\Omega_M /
 \Omega_E $ changes continuously from $1$ to about $1.27~(\simeq 4/\pi )$ (see 
Fig. 4), which is the average ratio of the Manhattan distance between two random
 points in a plane and the Euclidean distance between them [10, 5].

\noindent
{\bf 3 Conclusions}\\
\noindent
We note that the TSP 
on randomly diluted lattice is certainly a trivial problem when $p=1$ (lattice 
limit) as it reduces to the one-dimensional TSP (the connections in the optimal
 tour are between the nearest neighbours along the lattice).
Here $\Omega_E(p)=\Omega_M(p)=1$. 
 However, it is certainly NP- hard at the $p\rightarrow 0$ (continuum) limit,
where $\Omega_E \simeq 0.73$ and $\Omega_M \simeq 0.93$ (extrapolated for large 
system sizes $N$). We note that
$\Omega_M$ remains practically close to unity for all values of 
$p<1$. Our numerical results also suggest that $\Omega_M/\Omega_E\simeq 4/\pi$
as $p\rightarrow 0$.
It is clear that the problem crosses from triviality (for $p=1$) to the NP- hard
problem (for $p\rightarrow 0$) at a certain value of $p$. 
We did not find any irregularity in the variation of $\Omega$ at any $p$.
A naive expectation might be that around 
the percolation point, beyond which the marginally connected lattice spanning 
path is snapped off [7], the $\Omega_E$ or $\Omega_M$ suffers some irregularity.
The absence of any such irregularity can also be justified easily: the 
travelling salesman
has to visit all the occupied lattice sites (cities), not necessarily those on 
the spanning cluster. Also, the TSP on dilute lattices has got to accomodate 
the same kind of frustration as the (compact) self-avoiding chains on dilute 
(percolating) lattices, although there the (collapsed) polymer is confined only
 to the spanning cluster.   This indicates that the 
transition occurs either at $p=1_-$ or at $p=0_+$. From the consideration of 
frustration for the TSP even at $p=1_-$, it is almost certain that the 
transition occurs at $p=1$. 
However, this point requires further investigations.

\vskip 0.3in
\noindent
{\bf Acknowledgement} : We are grateful to O. C. Martin and A. Percus
for very useful comments and suggestions.

\newpage
\noindent
{\bf References}\\
\vskip .1in
\noindent
{\it e-mail addresses} :

\noindent
$^{(1)}$anirban@cmp.saha.ernet.in

\noindent
$^{(2)}$bikas@cmp.saha.ernet.in
\vskip .2 in

\noindent
1. M. R. Garey and D. S. Johnson, {\it Computers and Intractability: A Guide
to the Theory of NP- Completeness} (Freeman; San Franscisco) (1979).\\
\noindent
2. S. Kirkpatrick, C. D. Gelatt, Jr., and M. P. Vecchi, {\it Science},
{\bf 220}, 671 (1983).\\
\noindent
3. M. Mezard, G. Parisi and M. A. Virasoro, {\it Spin Glass Theory and Beyond}
(World Scientific; Singapore) (1987).\\
\noindent
4. J. Beardwood, J. H. Halton and J. M. Hammersley, {\it Proc. Camb. Phil. 
Soc.} {\bf 55}, 299 (1959); R. S. Armour and J. A. Wheeler, {\it Am. J. Phys.}
{\bf 51}, 405 (1983).\\ 
\noindent
5. A. Chakraborti and B. K. Chakrabarti, {\it cond-mat/0001069} (2000).\\ 
\noindent
6. A. Percus and O. C. Martin, {\it Phys. Rev. Lett.}, {\bf 76}, 1188 (1996).\\
\noindent
7. D. Stauffer and A. Aharony, {\it Introduction to Percolation Theory} (Taylor
and Francis; London) (1985).\\
\noindent
8. K. Barat and B. K. Chakrabarti, {\it Phys. Rep.}, {\bf 258}, 377 (1995).\\
\noindent
9. B. K. Chakrabarti, {\it J. Phys. A: Math. Gen.}, {\bf 19}, 1273 (1986).\\
\noindent
10. D. Dhar, M. Barma, B. K. Chakrabarti and A. Tarapder, 
{\it J. Phys. A: Math. Gen.}, {\bf 20}, 5289 (1987);
M. Ghosh, S. S. Manna and B. K. Chakrabarti,
{\it J. Phys. A: Math. Gen.}, {\bf 21}, 1483 (1988);
P. Sen and B. K. Chakrabarti,
{\it J. Phys. (Paris)}, {\bf 50}, 255, 1581 (1989).\\
\noindent
11. C. Hurtwitz, {\it GNU tsp\_ solve}, available at: http://www.cs.sunysb.edu/\~\\
\noindent
algorith/implement/tsp/implement.shtml\\

\newpage
\noindent
{\bf Figure captions}\\
\vskip .1 in
\noindent
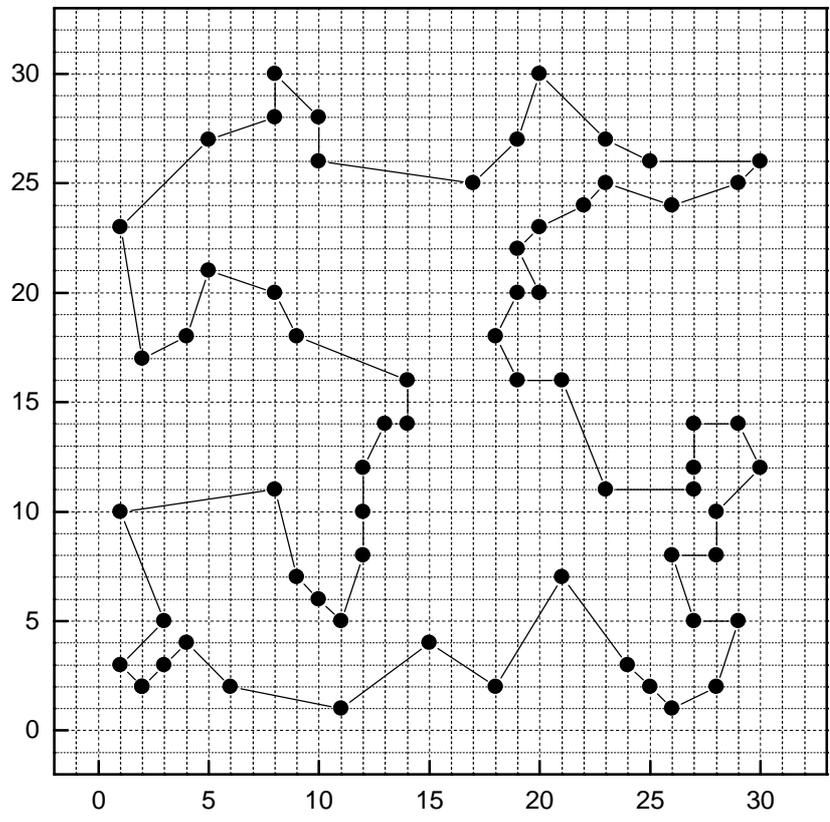
{\bf Fig. 1} : A typical TSP for ($N=)~64$ cities on a 
dilute lattice of size $L=30$. The cities are represented by black dots which 
are randomly occupied sites of the lattice with concentration $p=N/L^2\simeq 
0.07$. The optimized Euclidean path is indicated.\\
\noindent
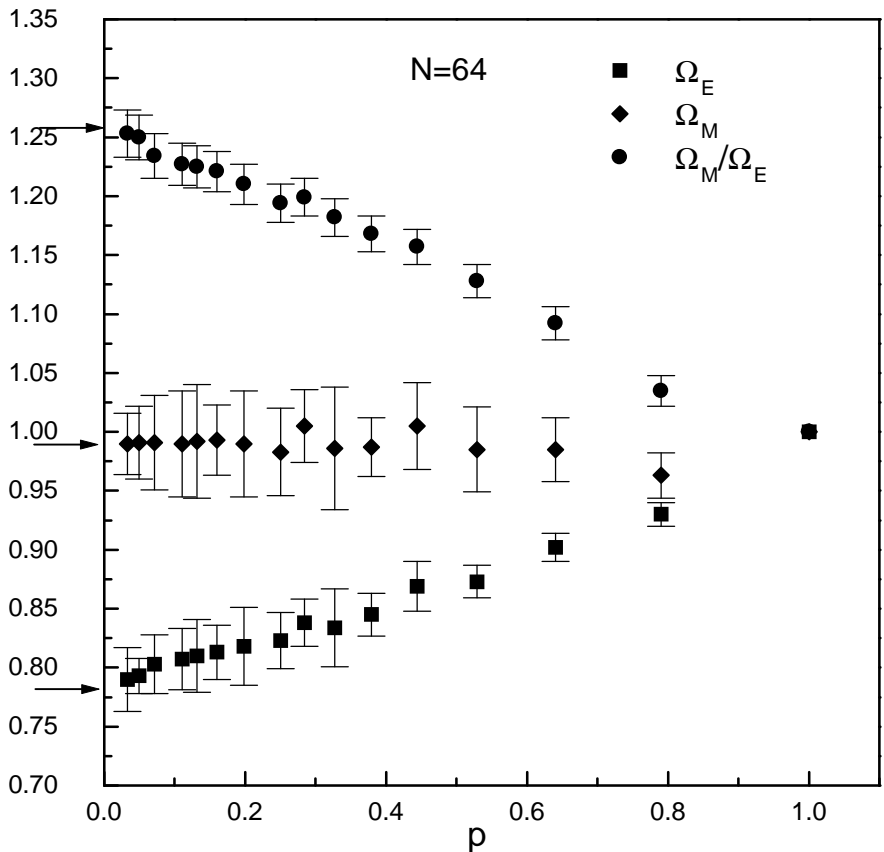
{\bf Fig. 2} : Plot of $\Omega_E$, $\Omega_M$ and $\Omega_M/\Omega_E$ against 
$p$ for $N=64$ cities, obtained using the optimization programs (exact).
 The error bars are due to configurational fluctuations.
The extrapolated values of $\Omega_E$, $\Omega_M$ and $\Omega_M/\Omega_E$ are
indicated by horizontal arrows on the y-axis.\\ 
\noindent
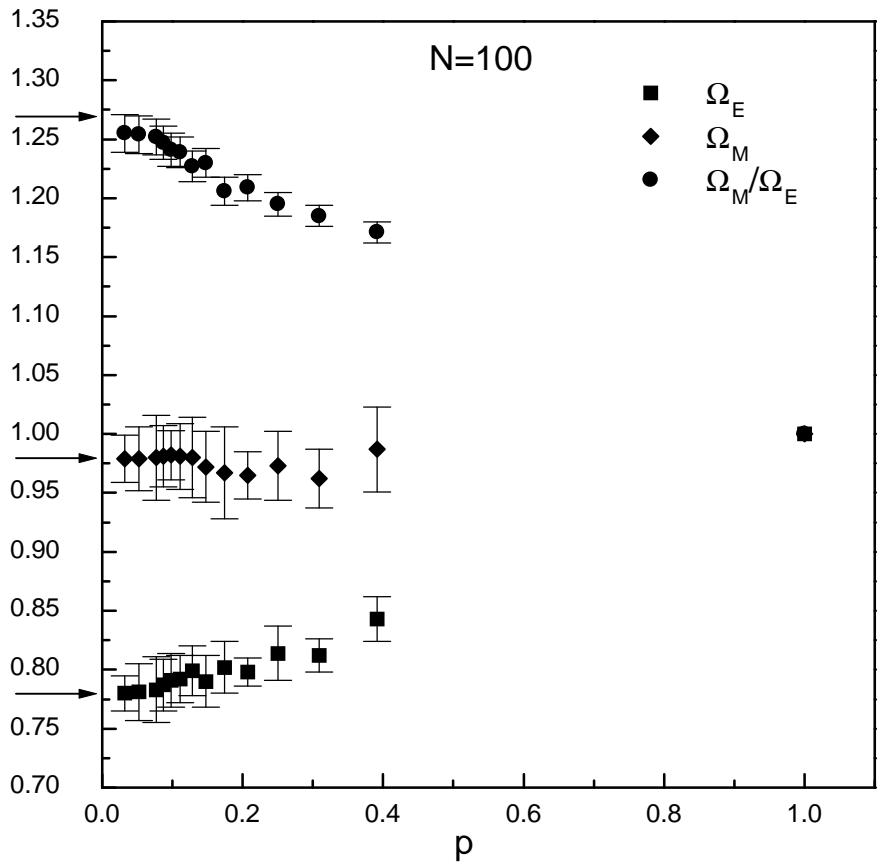
{\bf Fig. 3} : Plot of $\Omega_E$, $\Omega_M$ and $\Omega_M/\Omega_E$ against 
$p$ for $N=100$ cities, obtained using the optimization programs (exact).
 The error bars are due to configurational fluctuations.
The extrapolated values of $\Omega_E$, $\Omega_M$ and $\Omega_M/\Omega_E$ are
indicated by horizontal arrows on the y-axis.\\ 
\noindent
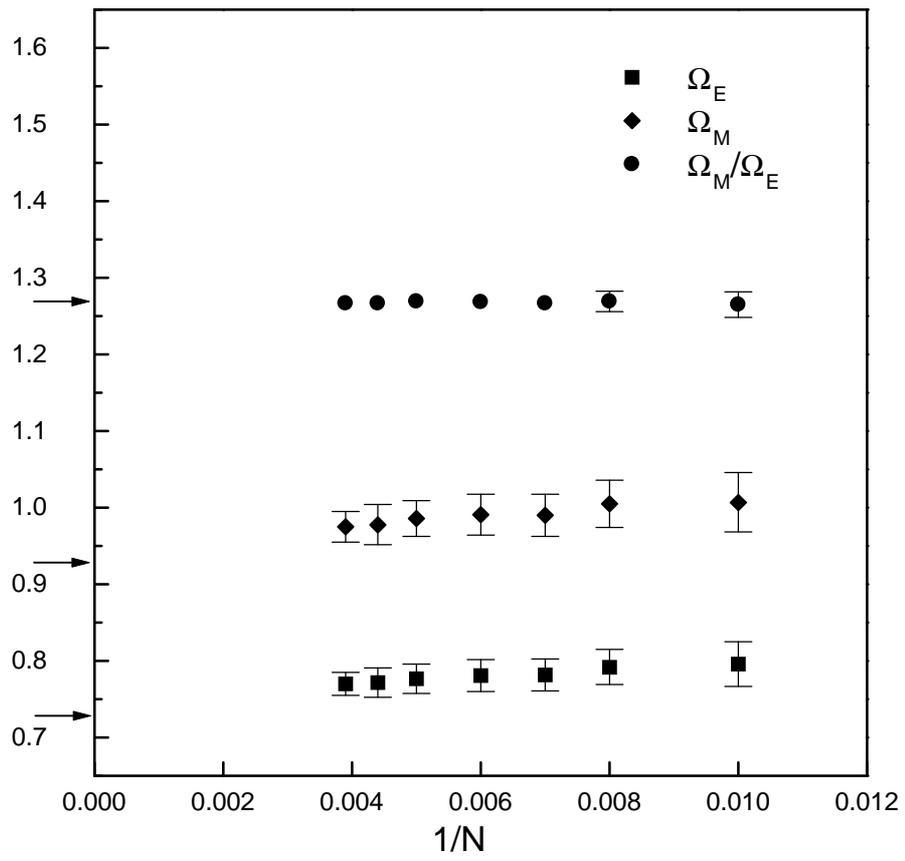
{\bf Fig. 4 } : Plots of $\Omega_E$ , $\Omega_M $ and of $\Omega_M/\Omega_E$ in
the $p\rightarrow 0$ limit, against $1/N$.
 The error bars are due to configurational fluctuations. 
The extrapolated value of $\Omega_E$ , $\Omega_M$ and $\Omega_M/\Omega_E$ in 
this $p\rightarrow 0$ limit for $N\rightarrow \infty$ are indicated by 
horizontal arrows on the y-axis.\\ 
 
\end{document}